\documentclass{pasj00}

\draft

\begin{document}
\SetRunningHead{S. Kato and J. Fukue}{Spin Estimate of BHBs by HF QPOs}
\Received{2005/00/00}
\Accepted{2006/00/00}

\title{Spin of Stellar-Mass Black Holes Estimated by 
a Model of Quasi-Periodic Oscillations}

\author{Shoji \textsc{Kato}}
\affil{Department of Informatics, Nara Sangyo University, Ikoma-gun,
       Nara 636-8503}
\email{kato@io.nara-su.ac.jp, kato@kusastro.kyoto-u.ac.jp}
\and
\author{Jun \textsc{Fukue}}
\affil{Astronomical Institute, Osaka Kyoiku University, 
         Asahigaoka, Kashiwara, Osaka, 582-8582}
\email{fukue@cc.osaka-kyoiku.ac.jp}

%

\KeyWords{accretion, accrection disks 
          --- quasi-periodic oscillations
          --- resonance
          --- stars: individual (GRO J1665-40, XTE J1550-564, GRS 1915+105)
          --- X-rays; stars} 

\maketitle

\begin{abstract}

We have proposed in previous papers that the high-frequency pair QPOs 
observed in black-hole binaries with frequency ratio 3:2 are inertial-acoustic oscillations (nearly horizontal oscillations with no node in the vertical direction) or g-mode 
oscillations, which are resonantly excited on warped relativistic disks.
The resonance occurs through horizontal motions.
In this model the dimensionless spin parameter $a_*$ of the central 
sources can be estimated when
their masses are known from other observations.
This estimate is done 
for three sources (GRO J1665-40, XTE J1550-564, GRS 1915+105).
For all of them we have $a_*\leq 0.45$.
 
\end{abstract}

\section{Introduction}

High-frequency quasi-periodic
oscillations (HF QPOs), whose frequencies are in the range of
100 to 450 Hz, have been observed in some black-hole binaries
and black-hole candidates.
One of characteristics of these  HF QPOs is that 
they often appear in a pair and their frequencies change little 
with time
\footnote
{
In the kHz QPOs of neutron-star X-ray binaries,
the frequencies (and their ratio) of the pair oscillations change 
with time.
This is a difinite difference between HF QPOs in black holes
binaries and kHz QPOs in neutron star binaries. 
In our warp models,  
kHz QPOs of neutron stars are interpreted as disk oscillations with
{\it vertical resonance} (see Kato 2005b), or as the case in which 
warp has precession (Kato 2005a).
}, 
keeping the frequency ratio close to 3:2.
These sources are GRO J1665-40 (300, 450 Hz), 
XTE J1550-564 (92, 184, 276 Hz) and GRS 1915+105 (41, 67, 113, 168 Hz)
(e.g., a review by McClintock and Remillard 2006).
Importance of commensurability of pair QPO frequencies on understanding
the disk structure in the innermost
region was emphasized by Abramowicz and Klu{\' z}niak (2001) and
Klu{\' z}niak and Abramowicz (2001).

It is known that  oscillations can be excited in a deformed disk
by resonant processes.
One well-known example is superhumps in tidally-deformed dwarf-novae
disks (Whitehurst 1988; Hirose, Osaki 1990; Lubow 1991).
Another example is a spiral pattern on ram-pressure
deformed galactic disks (Tosa 1994; Kato and Tosa 1994).
In black-hole X-ray binaries, similar types of resonant oscillations 
should occur when the disks are deformed.
We think that one of the most probable deformations
of disks in the innermost region is a warp.
Based on this idea, we examined excitation of
disk oscillations on warped disks (Kato 2003b, 2004b), 
and proposed a resonant excitation model of QPOs
(Kato 2004a,b, 2005a,b; Klu{\' z}niak et al. 2004).

In this warped-disk model, the high-frequency QPOs in black-hole 
binaries are g-mode oscillations or 
inertial-acoustic oscillations\footnote
{
In this paper inertial-acoustic oscillations represent  
the fundamental p-mode oscillations, in which oscillations are nearly 
horizontal and horizontal velocity has no node in the vertical direction.
In some recent papers by Kato, however, inertial-acoustic oscillations 
are treated together with g-mode oscillations, since 
in mathematical analyses of the present resonance problem 
both of them can be treated together in a pack without making distinction
(see Kato 2004b). 
Hence, when we used the term of g-mode oscillations, inertial-acoustic
oscillations were implicitly included there.
This was misleading.
Hence, in this paper we explicitly mention  
inertial-acoustic oscillations without including them in g-mode oscillations.
Here, it is noted that we use the following terminology for disk oscillations.
The modes in which horizontal velocity has no node in the vertical 
direction ($n=0$) is called inertial-acoustic oscillations (p-modes).
The modes with $n=1$ are g-modes and corrugation modes (c-modes).
The modes with $n\geq 2$ are g-modes and vertical p-modes}
or their combination, excited by a horizontal resonance.
If this resonance model is correct, it gives a way to estimate the spin 
of the central source from observed pair frequencies of QPOs, if 
the mass of the source is observationally known.

Recently, Shafee et al. (2006) evaluated the spins of two black-hole
sources (GRO J1655-40 and 4U 1543-47) whose masses are observationally 
known, by fitting their spectra with model spectra derived from current 
disk models.
Since one of the sources (i.e., GRO J1655-40) which they adopted 
has 3:2 pair frequencies, we can independently
estimate the spin of the source.
The purpose of this paper is to present the frequency-spin 
relation based on the warped-disk model and to estimate spins of
some black-hole X-ray binaries, including GRO J1655-40.

\section{Horizontal Resonances of G-Mode Oscillations and 
Inertial-Acoustic Oscillations in Warped Disks}

Here, we outline the essence of our resonance model in warped disks
[see figure 1 in Kato (2004a) and the similar figures in his subsequent
papers].
Let us consider a wave specified by ($\omega$, $m$, $n$),
where $\omega$ is the frequency of the wave, $m$ is the wavenumber
in the azimuthal direction, and $n$ is a number specifying the node
number in the vertical direction (for details, see, for example,  
Kato et al. 1998; Kato 2001).
A warp with no precession is described by (0, 1, 1),
since it is a kind of global, one-armed deformation.
Nonlinear interaction between the wave of ($\omega$, $m$, $n$) and the 
warp (1, 1, 0) produces oscillations specified by ($\omega$, $m\pm 1$, $n\pm 1$),
which are called here the intermediate oscillations.
If the amplitude of the wave mode of ($\omega$, $m$, $n$) is fixed,
the disk experiences forced oscillations due to the intermediate
oscillations.
The disk then resonantly responses to the intermediate oscillations
at some particular radius where the dispersion relation for these intermediate 
oscillations is satisfied.
At this radius energy exchange between the disk rotation
and the intermediate oscillations is realized. 
After this resonant inteaction, the intermediate oscillations feedback
to the original oscillation of ($\omega$, $m$, $n$) by making again 
nonlinear interaction with the warp.
This nonlinear feedback process amplifies or dampens the original
oscillation ($\omega$, $m$, $n$), since a resonant process is involved
in the feedback process (Kato 2003b, 2004b).

There are two kinds of resonances.
One is horizontal, and the other is vertical [see Kato 2004b for details].
Detailed examinations on resonant processes (Kato 2004b) show that 
the case in which resonance excites oscillations is the case in which
oscillations are inertial-acoustic oscillations and/or g-mode oscillations
and the resonance is horizontal.
Hence, in the followings, we restrict our attention only to the case. 

First, we remember that g-mode oscillations and inertial-acoustic 
oscillations with frequency $\omega$ 
and azimuthal wavenumber $m$ predominantly exist around the radius
specified by $(\omega-m\Omega)^2\sim \kappa^2$ by the following reasons.
Here, $\Omega$ and $\kappa$ are Keplerian and (radial) epicyclic 
frequencies, respectively.
In both cases of inertial-acoustic and g-mode waves, the group
velocity of these waves vanishes at the radius where 
$(\omega-m\Omega)^2=\kappa^2$.\footnote
{
In geometrically thin disks, the local dispersion relation of 
oscillations is given by
$$
    [(\omega-m\Omega)^2-\kappa^2][(\omega-m\Omega)^2-n\Omega_\bot^2]
    =c_{\rm s}^2k^2(\omega-m\Omega)^2,
                    \nonumber
$$
where $\Omega_\bot$, $c_{\rm s}$, and $k$ are, respectively,
the vertical epicyclic frequency, the acoustic speed, and the radial
wavenumber.
This dispersion relation gives the group velocity
($=\partial\omega/\partial k)$ as 
$$
   {\partial\omega\over\partial k}
    =\pm  c_{\rm s}{(\omega-m\Omega)^2[(\omega-m\Omega)^2-\kappa^2]^{1/2}
          [(\omega-m\Omega)^2-n\Omega_\bot^2]^{1/2}
      \over
      (\omega-m\Omega)^4-n\kappa^2\Omega_\bot^2}.   \nonumber
$$
}
That is, if we consider wave packets, they stay there for a long time
compared with in other places.
Hence, we think that the waves exist mainly around the radius
specified by\footnote{
The places of $(\omega-m\Omega)^2=\kappa^2$ are also particular places
in the sense that they are boundaries between the propagation
and evanescent regions of waves.
In the case of the inertial-acoustic waves, the propagation region
is described by $(\omega-m\Omega)^2>\kappa^2$, and the region of
$(\omega-m\Omega)^2<\kappa^2$ is the evanescent region.
In the case of the g-mode oscillations, the situation is changed.
That is, $(\omega-m\Omega)^2>\kappa^2$ is the evanescent region
and $(\omega-m\Omega)^2<\kappa^2$ is the propagation region.  
}
\begin{equation}
       (\omega-m\Omega)^2=\kappa^2.
\label{1}
\end{equation}

The nonlinear interaction of the above oscillations with 
a warp gives rise to intermediate oscillations
of  ($\omega$, $m\pm 1$).
These intermediate oscillations have resonant interaction with the disk
at the radii where the dispersion relation of the intermediate 
oscillations is satisfied (Kato 2003b, 2004b).
In the case of the horizontal resonances the radius is  
close to the radii specified by 
\begin{equation}
     [\omega-(m\pm 1)\Omega]^2=\kappa^2.
\label{2}
\end{equation}
It is important to note that the resonant radii are 
independent of the vertical structure of the oscillations,
i.e., independent of $n$.
 
The resonant radii and the radii where the oscillations
predominantly exist must be the same for resonant interactions
to occur efficiently.
That is, equations (\ref{1}) and (\ref{2}) must be satisfied 
simultaneously, which gives
\begin{equation}
      \kappa={\Omega \over 2}.
\label{3}
\end{equation}
This is the condition determining the resonant radius.
From equation (\ref{1}), we then see that the frequencies of 
resonant oscillations  are $m\Omega\pm \kappa$ at the resonant radius.
The above argument is free from the metric.
That is, the above resonant condition is valid  
even in the case of the Kerr metric, if 
the angular velocity of the Keplerian rotation,
$\Omega$, and the epicyclic frequency, $\kappa$, in the
Kerr metric are adopted.

\section{Resonant Radius and Frequencies of Resonant Oscillations}

In the limit of non-rotating central source 
(i.e., the metric is the Schwarzschild one),
the condition, $\kappa=\Omega/2$, is realized at $4.0r_{\rm g}$, 
which is just the radius where $\kappa$ becomes the maximum.
Here, $r_{\rm g}$ is the Schwarzschild radius defined by 
$r_{\rm g}=2GM/c^2$, $M$ being the mass of the central source.
As the spin parameter, $a_*$, increases the resonant radius, $r_{\rm c}$,
decreases.
The $r_{\rm c}$ -- $a_*$ relation derived from the resonant condition, $\kappa=\Omega/2$, is shown in figure 1.

Next, we calculate frequencies of inertial-acoustic and/or g-mode
oscillations which
have resonance at $\kappa=\Omega/2$.
As mentioned before, they are $m\Omega\pm \kappa$ at the
resonant radius.
They are a set of frequencies, since there are various $m$.
Among them the most observable ones will be those with
small number of $m$.
The axially symmetric oscillations, $m=0$, however, will be less
observable by the very nature of symmetry.
Hence, the oscillations which will be most interesting in relation to
observed QPO frequencies are those with $m=1$ or $m=2$.
Considering this situation, we introduce, for convenience,
symbols given by
\begin{equation}
    \omega_{\rm H}=(\Omega+\kappa)_{\rm c}, \quad 
    \omega_{\rm L}=(2\Omega-\kappa)_{\rm c}, \quad 
    \omega_{\rm LL}=(\Omega-\kappa)_{\rm c},
\label{4}
\end{equation}
where the subscript c denotes the values at the resonant radius, 
$\kappa=\Omega/2$.
It is noted that $\omega_{\rm H}$ and $\omega_{\rm L}$ are equal,
i.e., $\omega_{\rm H}=\omega_{\rm L}$.
Outside the resonant radius (i.e., $r>r_{\rm c}$), $\Omega+\kappa$ 
is larger than
$2\Omega-\kappa$ since $\kappa>\Omega/2$ there.
Inside the resonant radius, $2\Omega-\kappa$ is larger than 
$\Omega+\kappa$.

\subsection{Source State and QPO Frequencies}

The next problem to be examined is the relation between the frequencies of
disk oscillations mentioned above and the observed QPOs frequencies. 
For simplicity, let us neglect effects of disk rotation (such as Doppler
boosting) and geometrical effects (such as gravitational bending of light rays
and occultation).
Then, no luminosity variation is observed in  
geometrically thin, no warped disks, even  if  rotating non-axisymmetric
oscillations are superposed on the disks.
This means that some careful consideration on geometrical states of disks 
is necessary.
Observations show (Remillard 2005) that all high-frequency QPOs are 
associated with the steep power-law state of sources.
They are not observed in the thermal state (i.e., the soft/high
state with no corona), nor in the hard state (i.e., hard/low state
with no thermal disk component).
It is noted that in the steep power-law state a compact hot torus
(corona) and a thermal disk coexist in the innermost region.
Observations further show that the QPOs are observed in the 
high energy photons of the power-law
component, not in the soft photons of the thermal disk component.

This observational evidence suggests that a thermal disk is necessary
as a place where oscillations are generated, but the observed QPO 
photons are those Comptonized in the hot compact corona (a hot torus).
If this picture is adopted,  one-armed oscillations 
are observed as time variations with twofold frequency, as
described below.

Let us consider one-armed disk oscillations propagating in the 
azimuthal direction with angular frequency $\omega$.
The hot disk region associated with the disk oscillations
is assumed to be inside a torus.
Now,  we consider the path of observed photons which are originally emitted 
from the hot region of the disk as soft photons and are observed as 
high energy photons by Comptonization in the torus.
The path length of the photons in the torus dependes on the phase
relation between the hot region and the observer, as shown in figures
2 and 3.
In the phase shown in figure 2, the path length of photons in torus is
short.
(This phase is called hereafter phase 0.)
In the phase shown in figure 3, however, the path length within the 
torus is long.
The latter occurs when the phase is close to 0.75 as well as 0.25.
In the phase 0.5 the path in the torus is shorter than that in the phase of figure 3,
but longer than that in the phase of figure 2 (phase 0).
Hence, observed Comptonized photon numbers will vary as shown in
figure 4.
That is, we have  two peaks during one cycle of the oscillations.


Here, a brief comment is made on depths of the primary
minimum (phase 0) and secondary minimum (phase 0.5) in figure 4.
In the phase of the primary minimum, the path length of photons in the torus is short, but they pass through an inner hot and dense region of the torus (see figure 2).
This will increase the Comptonized photon flux, compared with that in the case in which photons 
pass an outer cool and less dense region.
The phase of the secondary minimum (phase 0.5) correspons to the latter
case.
This consideration suggests that the difference between the Comptonized photon fluxes in phases 0 and 0.5 is smaller than that simply estimated from 
the difference of geometrical path lengths. 
Figure 4 should be regarded as results in which the above effects are already taken into account.

Figure 4 shows that one-armed oscillations with frequency $\omega$ bring
about two time-varying components with  $\omega$ and $2\omega$. 
Let us now roughly estimate the amplitude ratio of the two components 
from the light curve in figure 4.
The flux, $f(t)$, shown in figure 4 will be approximated by
\begin{equation}
         f(t) =(1+A)-{\rm cos}(4\pi t)-A{\rm cos}(2\pi t),
\end{equation}
with $A$ moderately smaller than unity, where $t$ represents the phase of 
the light curve (i.e., $t=0$ at phase 0 and $t=1$ at phase 1).
The amplitude of the $2\omega$-oscillation is normalized to unity and that of
the $\omega$-oscillation is $A$.
The total flux is normalized to become zero at phase 0.
Then, the maximum of the flux is realized near $t=1/4$ and $3/4$ as far as $A$ is 
moderately smaller than unity, and is about $2+A$.
The flux at the secondary minimum ($t=0.5$) is $2A$.
That is, the flux ratio of the secondary minimum to the maximum is
roughly $2A/(2+A)$.
The case where the secondary minimum is as deep as the primary minimum,
i.e., $2A/(2+A)=0$, is realized if $A=0$.
That is, in this case we have only $2\omega$ oscillation, as expected.
If, for example, the flux ratio is 0.3, the amplitude ratio is found to be roughly
0.35.
That is, the amplitude of the oscillation with $\omega$ is smaller than that with $2\omega$ by about factor 3.


In the case in which the observer is in a direction close to the
edge-on, a light ray leaving the torus to go to the observer 
may enter again into another part of the torus on the way of the path.
Furthermore, the Doppler effects are not negligible on light variation.
Such cases of high inclination angle, however, will not be
the major cases in which QPOs are observed, since outer 
parts of disk will screen QPO photons from the observer.

In the case of two-armed oscillations, we can easily find that the main
frequency of observed luminosity variation is the same as that of the
oscillations. 

These considerations suggest that the resonant oscillations with 
frequency $\omega_{\rm LL}$ mainly give rise to QPOs whose frequencies are
$2\omega_{\rm LL}$, since they are one-armed oscillations, i.e., $m=1$.
Hence, we think that the observed main 
frequencies of QPOs, i.e., the frequencies of the pair QPOs, are  
$\omega_{\rm L}(=\omega_{\rm H}$) and $2\omega_{\rm LL}$.
Their frequency ratio 
is just 3:2, i.e.,
\begin{equation}
   \omega_{\rm L}(=\omega_{\rm H}) : 2\omega_{\rm LL}=3:2.
\end{equation}
In the present disk-oscillation model there is no reason 
why oscillations with $\omega_{\rm LL}$ are not observed, although
their amplitude may be small.
We think that these oscillations are really observed in some sources.
In XTE J1550-564 three QPOs are observed whose frequencies are
276Hz, 184Hz, and 92Hz. 
Their frequency ratios are just 3:2:1, suggesting that $\omega_{\rm LL}$
has been observed.
Furthermore, in a black-hole X-ray transient XTE J1650-500, QPO
frequencies vary with time, but their frequencies are consistent with
being 1:2:3 harmonics (Homan et al. 2003), suggesting that 
$\omega_{\rm LL}$ has been also observed in this source.

One may think why QPOs with frequency $2\omega_{\rm H}$ are not
observed. (It is noted that the oscillations with $\omega_{\rm H}$ are
one-armed.)
We think that they should be observed, but there is still no serious 
attempt to detect such high frequency QPOs, since
the frequency is higher than the Keplerian frequency in the innermost 
region of disks.   

\section{Estimate of Spin from Pair QPO Frequencies}

In the case in which the central source is non-rotating, the resonance
occurs at $4.0r_{\rm g}$, and $\omega_{\rm L}(=\omega_{\rm H})$ can 
be easily expressed as
\begin{equation}
    \omega_{\rm L}=2.14\times 10^3\biggr({M\over M_\odot}\biggr)^{-1}
       \ {\rm Hz}.
       \qquad (a_*=0)
\label{5}
\end{equation}
Masses of three sources (GRO J1665-40, XTE J1550-564, GRS 1915+105)
which display a pair of HF QPOs have been obtained from spectroscopic 
observations.
Using the data, McClintock and Remillard (2005)
derived an interpolation formula giving a relation between observed 
frequencies of HF QPOs and $M$, which is
\begin{equation}
   3\nu_0 = 2.79\times 10^3\biggr({M\over M_\odot}\biggr)^{-1}
       \ {\rm Hz},
\label{6}
\end{equation}
where $\nu_0$ is the fundamental frequency of 3:2:1, and thus $3\nu_0$
corresponds to $\omega_{\rm L}$ in our model.
The frequency $\omega_{\rm L}$ for $a_*=0$ is smaller than $3\nu_0$,
suggesting that the central sources are certainly rotating.

The dependence of $\omega_{\rm L}$ on the spin parameter $a_*$ is 
numerically obtained by substituting $r_{\rm c}$ obtained by solving 
equation (\ref{3}) into the expression for $\omega_{\rm L}$ [equation 
(\ref{4})].
The results are shown in figure 5.
For the three sources, where $M$ and $3\nu_0$ are known, 
the spin parameter $a_*$ can be calculated, assuming that
the observed $3\nu_0$ is $\omega_{\rm L}(=\omega_{\rm H})$.
The results are shown in table 1 (see also table 3 of Kato 2004b).
As shown in table 1, the value of spin parameter $a_*$ derived 
for GRO J1665-40 is $a_*=0.31$ -- $0.42$, which is somewhat smaller than 
$a_*=0.65$ -- $0.75$ derived by Shafee et al. (2006)  from a spectrum 
fitting.  
In the case of GRS 1915+105, the value of $a_*$ is negative if
$M\sim 10.0M_\odot$ is adopted.
This suggests that the mass is much larger than $10M_\odot$, closer 
to $18.0M_\odot$.

 
\begin{table}
  \caption{Estimated spin parameter $a_*$.}\label{tab:first}
 \begin{center}
  \begin{tabular}{cccc}
\hline\hline
 Sources & $3\nu_0$(Hz)  &  $M/M_\odot$  & $a_*$  \\
\hline
 GRS 1915+105  &  168  & 10.0 -- 18.0 & negative -- 0.44 \\
 XTE 1550-564  &  276  & 8.4 -- 10.8  &  0.11 -- 0.42  \\
 GRO 1655-40   &  450  & 6.0 -- 6.6 & 0.31 -- 0.42  \\
\hline
 \end{tabular}
  \end{center}
\end{table}

\section{Propagation Regions}

It is worthwhile to note that the propagation region 
of inertial-acoustic oscillations and that of g-mode ones 
are different, even when their frequencies are the same.
They are inside or outside of the resonant radius, depending on the
modes.
The propagation regions of the inertial-acoustic oscillations with
frequency $\omega$ and azimuthal wavenumber $m$ are in the region
described by $(\omega-m\Omega)^2>\kappa^2$, which is 
$\omega>m\Omega+\kappa$ or $\omega<m\Omega-\kappa$.
In the case of g-mode oscillations, the region is $(\omega-m\Omega)^2
<\kappa^2$, which is $m\Omega-\kappa < \omega <m\Omega+\kappa$.
To demonstrate these situations, we show in figure 6 the propagation 
regions of inertial-acoustic oscillations and those of g-mode oscillations 
whose frequencies are $\omega_{\rm H}$, $\omega_{\rm L}$, 
and $\omega_{\rm LL}$.

As shown in figure 6 and mentioned above, the propagation regions of 
inertial-acoustic
oscillations and those of g-mode oscillations are in the opposite sides
of the resonant radius, when their frequencies are the same.
In the propagation regions of the g-mode oscillations, there is 
a corotation radius, i.e., the radius where $\omega-m\Omega=0$.
At the corotation radius the g-mode oscillations are damped (Kato 2003a,
Li et al. 2003).
The inertial-acoustic oscillations which propagate inward from the
resonant radius will be partially reflected back near the inner edge of
the disk, which may lead to quasi-trapped oscillations.
These considerations may suggest that the main contributor to
the HF QPO may be inertial-acoustic oscillations, rather than g-mode 
oscillations.
 
\section{Discussion}
 
The basic idea of our model is that the high-freuqency 
QPOs are disk oscillations and a deformation of the disk is the essential
cause of their excitation.
As the cause of disk deformation we consider warp.
This is because warp will be one of the most conceivable deformation of
disks in the innermost region. 
As mentioned before, 
the QPOs are  associated 
with the steep power-law (SPL) state and are certainly not
in the thermal state where the disk consists only of a thermal disk
component (Remillard 2005).
In the SPL state a compact corona and a thermal disk coexist.
We suppose that triggeres forming a compact high-temperature torus 
in the innermost region will not generally axisymmetric since the disks are
highly turbulent,  and deform the
disk as well as formation of a torus.
This will be one of possible causes of formation of warped disk.


Let us denote the observed upper and the lower frequencies of the pair
QPOs by $\nu_{\rm u}$ and $\nu_{\rm l}$.
Then, as mentioned before, our present model predicts the presence
of QPOs with frequency of $2\nu_{\rm u}$.
Analysis of obervational data to see whether QPOs of $2\nu_{\rm u}$
are present or not is a cruical check of the present model.

\bigskip
The author thanks the referees and S. Mineshige for valuable commnents.

\bigskip
\leftskip=20pt
\parindent=-20pt
\par
{\bf References}
\par
Abramowicz, M. A., \& Klu{\' z}niak, W. 2001, A\&A, 374, L19 \par
Hirose, M., Osaki, Y. 1990, PASJ, 42, 135\par
Homan, J. Klein-Wolt, M., Rossi, S. Miller, J.M., Wijnands, R., Belloni, 
T., van der Klis, M., Lewin, W.H.G., 2003, ApJ, 586, 1262 \par
Kato, S. 2001, PASJ, 53, 1\par 
Kato, S. 2003a, PASJ, 55, 257 \par
Kato, S. 2003b, PASJ, 55, 801\par
Kato, S. 2004a, PASJ, 56, 559 \par
Kato, S. 2004b, PASJ, 56, 905\par
Kato, S. 2005a, PASJ, 57, L17 \par
Kato, S. 2005b, PASJ, 57, 699 \par
Kato, S., Fukue, J., \& Mineshige, S. 1998, Black-Hole Accretion Disks 
  (Kyoto: Kyoto University Press)\par
Kato, S., Tosa, M. 1994, PASJ, 46, 559 \par
Klu{\' z}niak, W., \& Abramowicz, M. 2001, Acta Phys. Pol. B32, 3605   \par
Klu{\' z}niak, W., Abramowicz, M. A., Kato, S., Lee, W. H., \& Stergioulas,
   N. 2004, ApJ, 603, L89 \par 
Li, L.-X., Goodman, J., Narayan, R. 2003, ApJ, 593,980 \par
Lubow, S.H. 1991, ApJ, 381, 259\par
McClintock, J.E., Remillard, R.A. 2005, "Black Hole Binaries", in
   Compact Stellar X-ray Sources, eds. W.H.G. Lewin and M. van der Klis,
   Cambridge University Press, Cambridge, in press; astro-ph/0306213 \par
Remillard, R.A. 2005, Astron. Nachr. vol. 326; astro-ph/0510699 \par 
Shafee, R., McClintock, J.E., Narayan, R., Davis, S.W., Li, L.-X.,
    Remillard, R.A. 2006, ApJ, 636, L113; astro-ph/0508302\par     
Tosa, M. 1994, ApJ, 426, L81 \par
Whitehurst, R. 1988, MNRAS 232, 35    \par
\bigskip\bigskip

\begin{figure}
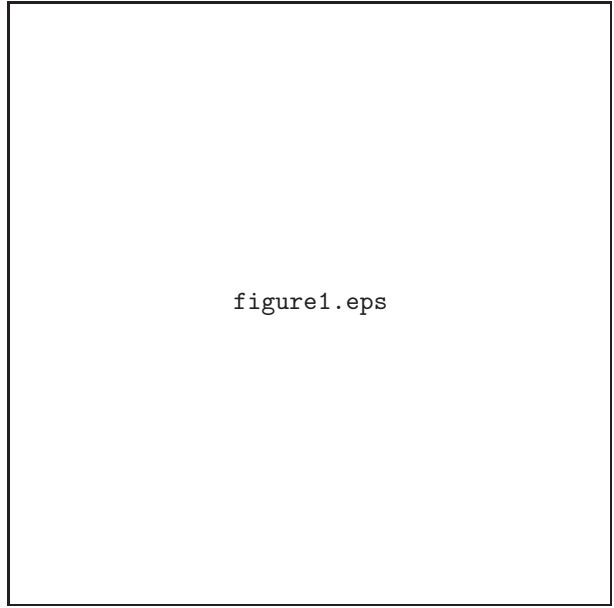

  \begin{center}
    \FigureFile(80mm,80mm){figure1.eps}
  \end{center}
  \caption{
     Resonant radius as a function of the spin parameter $a_*$.    
     The resonant radius is given by $\kappa=\Omega/2$.}
\label{fig:figure 1}
\end{figure}

\begin{figure}
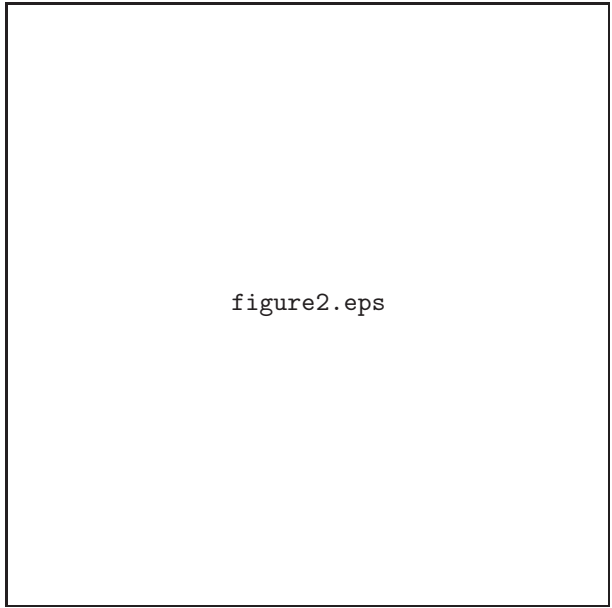

  \begin{center}
    \FigureFile(80mm,80mm){figure2.eps}
  \end{center}
  \caption{
     A schematic picture showing the light path from a hot region in disks 
    to an observer in the phase in which the hot region is just in the opposite
side of the central source to the observer.
The path within the torus is shown by dashed line.
It is noticed that the path length in the torus is short, and the observed QPO
photons are not many.
This phase is referred to phase 0, and the phase in which the hot region of the disk
is between the central source and the observer is phase 0.5 }
\label{fig:figure 2}
\end{figure}

\begin{figure}
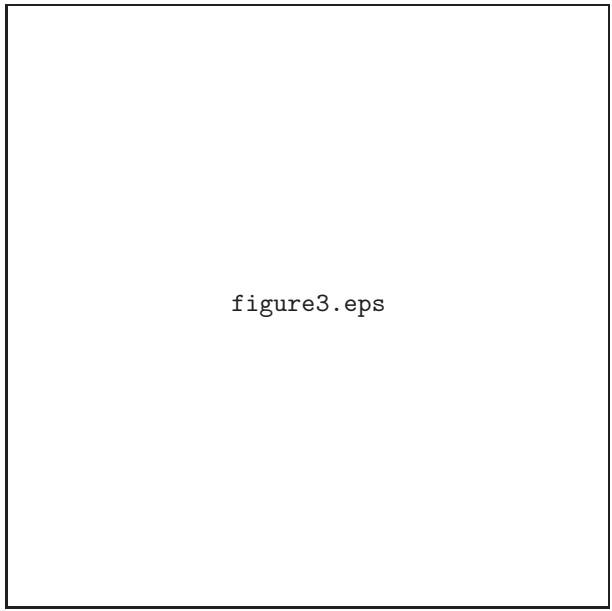

  \begin{center}
    \FigureFile(80mm,80mm){figure3.eps}
  \end{center}
  \caption{
    A schematic picture showing a straight light path from the hot region in disks
to an observer in a phase close to 3/4.
The part of the pass within the torus is shown by dashed line.
The pass within the torus is the longest in this phase as well as in a phase close to
1/4, compared with in other phases.
 }
\label{fig:figure 4}
\end{figure}

\begin{figure}
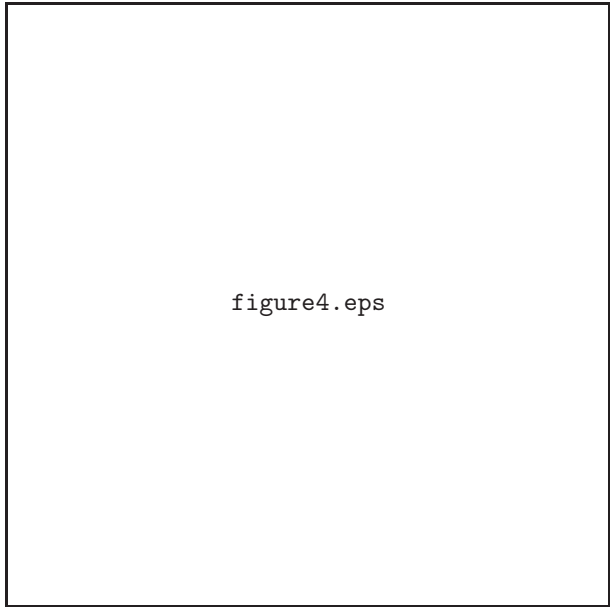

  \begin{center}
    \FigureFile(80mm,80mm){figure4.eps}
  \end{center}
\caption{
   A schematical light-curve during one revolution
of an one-armed oscillation around a central source.
We have two peaks during one cycle of oscillations around phases of
0.25 and 0.75.
}
\end{figure} 

\begin{figure}
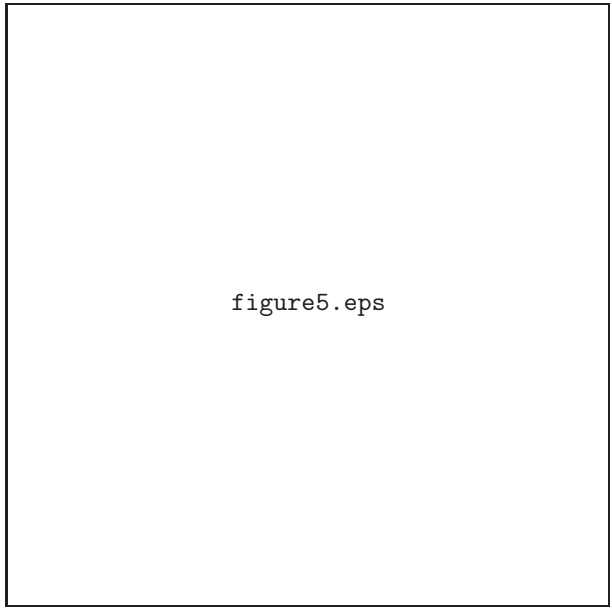

  \begin{center}
    \FigureFile(80mm,80mm){figure5.eps}
  \end{center}
  \caption{
      The frequency $\omega_{\rm L}$ of the upper HF QPO as a 
      function of the spin parameter $a_*$.} 
\label{fig:figure 5}
\end{figure}

\begin{figure}
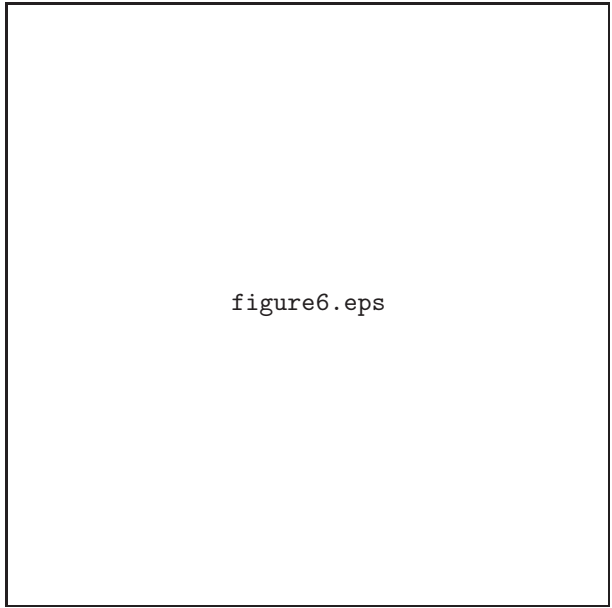

  \begin{center}
    \FigureFile(80mm,80mm){figure6.eps}
  \end{center}
  \caption{
     Propagation regions of the resonant oscillations whose frequencies are 
    $\omega_{\rm H}$, $\omega_{\rm L}$, and $\omega_{\rm LL}$.
    For each, the region, which is shown by arrow, depends on whether 
    the oscillations are
    inertial-acoustic modes or g-modes.
    The symbols attached to arrow show frequency and mode.
    For example, $\omega_{\rm H,p}$ denotes the inertial-acoustic
    oscillations of resonant frequency $\omega_{\rm H}$.
    The case of g-mode oscillations, the subscript g is attached
    instead of p.
    To make clear the propagation regions, 
    curves representing radial distributions of $\kappa$, $\Omega$, 
    $\Omega\pm \kappa$, $2\Omega$ and $2\Omega\pm \kappa$ are shown.
    The vertical line shows the radius where the resonance occurs.
    This figure is drawn for $M=10M_\odot$ and $a_*=0.2$.
} 
\label{fig:figure 6}
\end{figure}

\end{document}